\documentstyle[epsf]{article}

\begin{document}

\title{How effective is advertising in duopoly markets?}

\author{K. Sznajd--Weron\\
Institute of Theoretical Physics, University of Wroc{\l}aw, \\
pl. Maxa Borna 9, 50-204 Wroc{\l}aw, Poland\\
E-mail: kweron@ift.uni.wroc.pl\\
\\
R. Weron\\
Hugo Steinhaus Center for Stochastic Methods,\\
Wroc{\l}aw University of Technology, 50-370 Wroc{\l}aw, Poland\\
E-mail: rweron@im.pwr.wroc.pl
}

\maketitle

\begin{abstract}
A simple Ising spin model which can describe the mechanism of advertising
in a duopoly market is proposed. In contrast to other agent-based models, the influence
does not flow inward from the surrounding neighbors to the center site, but spreads
outward from the center to the neighbors. The model thus describes the spread of opinions
among customers.
It is shown via standard Monte Carlo simulations that very simple rules and inclusion 
of an external field -- an advertising campaign -- lead to phase transitions.
\end{abstract}

\noindent
{\bf Keywords:}
econophysics, advertising, oligopoly, duopoly, Ising model, agent-based model

\noindent
{\bf PACS:} 05.50.+q, 05.10.Ln, 89.20.-a, 89.90.+n

\vspace{1cm}

\section{Introduction}

Oligopoly is defined as a market structure where there are normally a few large sellers.
Duopoly is a special case of oligopoly, where only two firms compete with each other \cite{friedman83,tirole88}. Actually there are a number of market possibilities that could be called oligopoly (or duopoly): a market with a few, large manufacturers selling the same product (e.g. power producers, electricity utilities \cite{wolfram99}, steel or oil refining industries); a market with a few, large producers selling a differentiated product (e.g. baby food, beer, cereal or automobile industries); or even a market with one large seller and some smaller sellers (e.g. aircraft (Boeing), soup (Campbell), film (Kodak) or detergent (Procter \& Gamble) industries). If fact, in most countries markets are dominated by one or a handful of firms. This is normal and expected behavior \cite{hsw97,cb98}. 

What distinguishes oligopoly (or duopoly) from the other forms of market structure is that the firms in an oligopolistic market must be aware of and be judging the reaction of other firms on the market \cite{mt01}. With a finite number of sellers there is often intense competition with respect to price, quality, and quantity produced. This is where advertising enters the picture. 

Consumers need information about product characteristics and prices to make rational (efficient) decisions. Advertising can be a low-cost means of providing it. Moreover, by supplying information about the various competing goods that are available, advertising diminishes monopoly power. In fact, advertising is frequently associated with the introduction of new products designed to compete with existing brands \cite{cb98}. Could Toyota and Honda have so strongly challenged U.S. auto producers without advertising? Could Federal Express have sliced market share away from UPS and the U.S. Postal Service without advertising? How about Dialog, an upstart private telecom in southern Poland, which recently has gained market share from long-time monopolist Telekomunikacja Polska?

When is advertising effective and when is it not? This question has often stimulated heated debate in the world of marketing and advertising \cite{barry87,scholten96}. In this paper we try to answer the above question via Monte Carlo simulations of a simple, two-dimensional Ising spin model. 

An Ising spin system, probably the most often used model in statistical physics, is a system with binary variables. These variables have been successfully interpreted as occupied/empty sites in percolation theory \cite{stauffer85}, up/down steps in DNA walk \cite{peng_etal92,stanley_etal99}, healthy/sick individuals in genetics \cite{oos99}, bulls/bears in financial markets \cite{cb00} or government/opposition in political models \cite{galam00}. In our case they correspond to product A/product B customers.

\section{The model}

We model a generic duopoly market by a two-dimensional lattice $L \times L$ with periodic boundary conditions. Each site of the lattice is "occupied" by an individual (a customer), who is characterized by a spin, S, that represents his/her support for one of the two equivalent products in the market. The spin can be either an up-spin (customer of product A, e.g. Dialog telecom, denoted by $\uparrow$) or a down-spin (customer of product B, e.g. long-time monopolist Telekomunikacja Polska, denoted by $\downarrow$). 

For a given set of rules, governing the behavior of the customers, we study the system via Monte Carlo simulations. In contrast to usual majority rules \cite{adler91}, in our model the influence does not flow inward from the surrounding neighbors to the center site(s), but spreads outward from the center to the neighbors. The model thus describes the spread of opinions. This approach is borrowed from the original one-dimensional Sznajd model \cite{s-ws00} (for a review see \cite{stauffer01,schechter02}), where two people forming a nearest-neighbor bond and sharing the same opinion convinced their neighbors of this opinion. 
In the two-dimensional setup it is natural to select a "panel" of four neighboring sites which may influence their eight nearest neighbors (see Fig. 1), however, other possibilities have been also studied \cite{sso00,ochrombel01}.
In our simulations we use random sequential updating, i.e. in each simulation step one of the $L \times L$ customers (denoted by an asterisk "$*$" in Fig. 1(a)) in the square lattice is selected randomly and then together with his/her three neighbors  -- right, lower right and lower -- forms a panel. 

\begin{figure}[htbp]
\centerline{(a) \epsfxsize=3cm \epsfbox{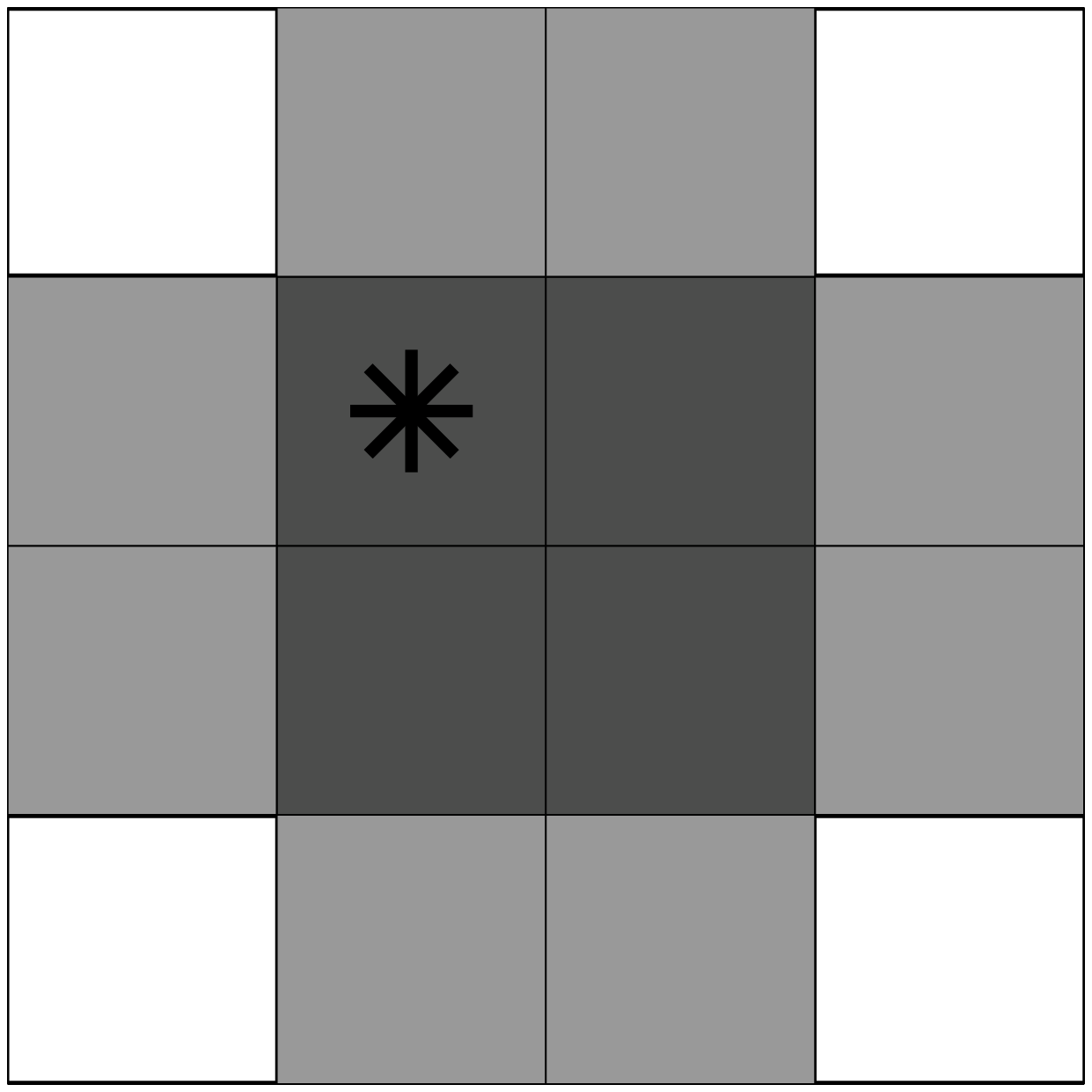} \quad
            (b) \epsfxsize=3cm \epsfbox{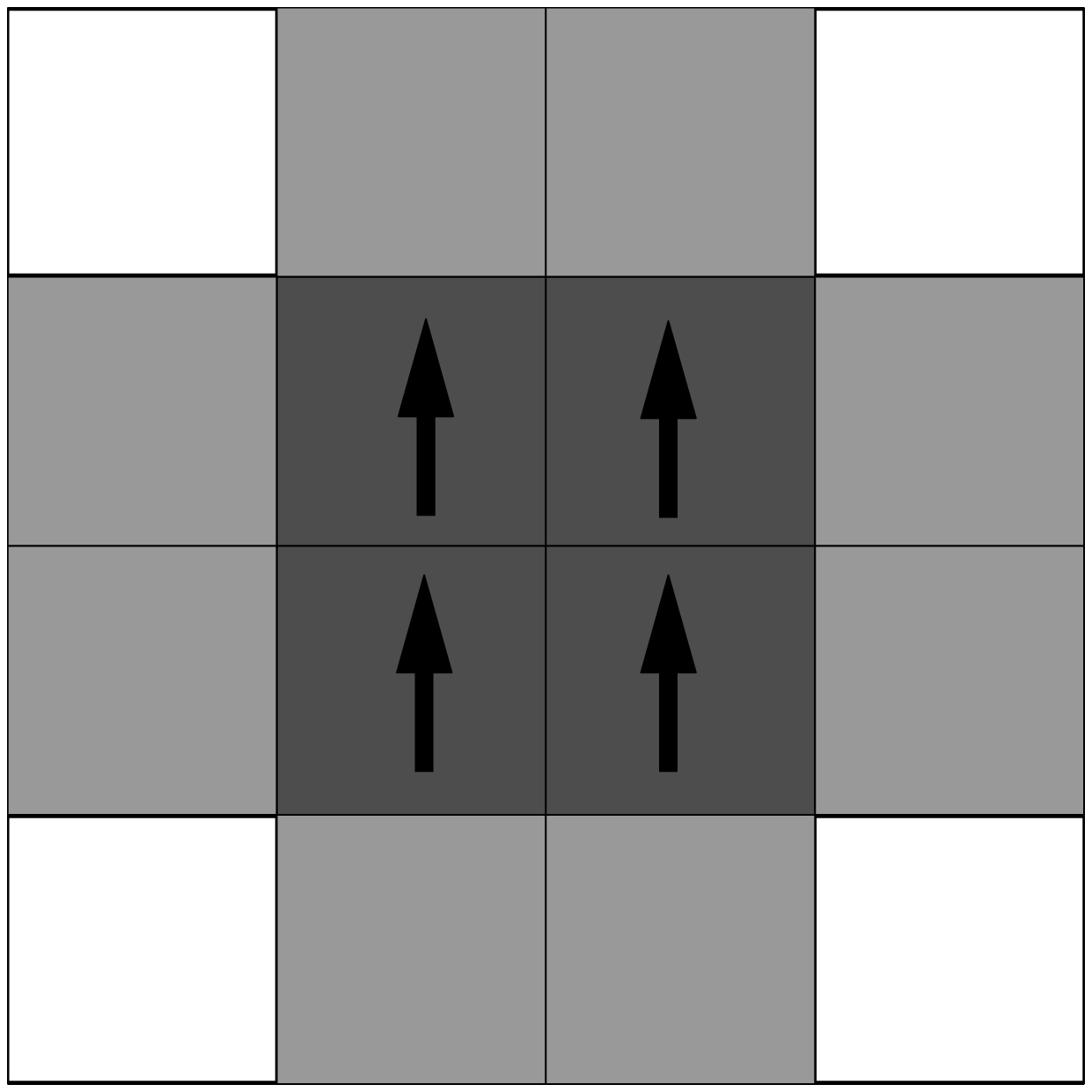} \quad
            (c) \epsfxsize=3cm \epsfbox{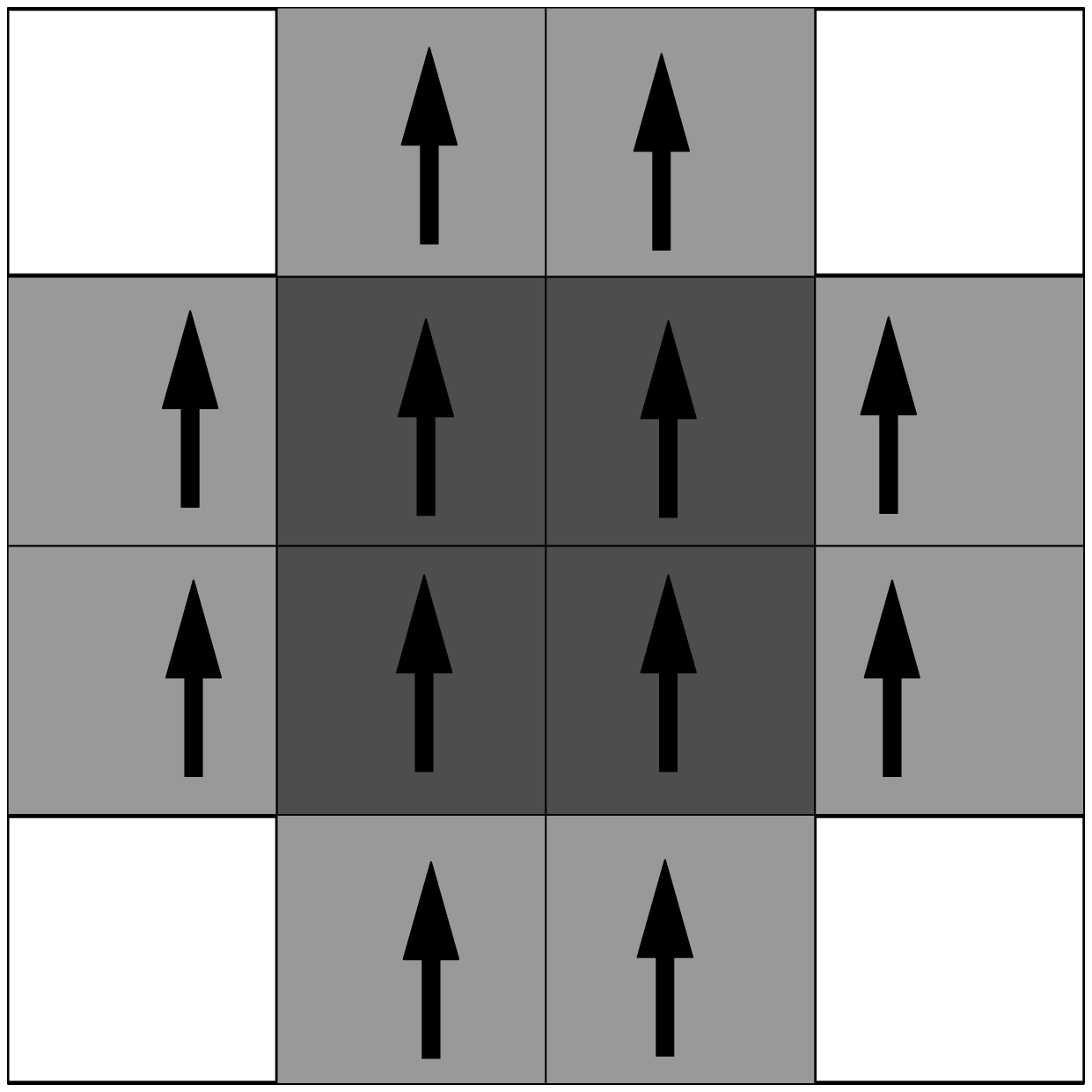}}
\caption{
(a) A customer (denoted by $*$) is selected randomly and then together with his/her three neighbors forms a panel (dark gray sites) which can influence eight nearest neighbors (light gray sites). 
(b) If a panel consists of four customers sharing the same opinion (spins in the same direction, here up-spins) then (c) all eight neighbors will turn in the same direction.}
\end{figure}

The first rule we impose on the system is the following. If all four individuals in a panel support the same product (the sum of spins $\sum_{i=1}^4 S_i$ is equal to four) they will convince all eight of their nearest neighbors resulting in changing their orientation in the direction of the spins in the panel, see Fig. 1.
If one of the spins in the panel has the opposite orientation to the other three spins then the neighbors change their orientation to the orientation of the majority with probability $3/4$. In the case when there is no majority, i.e. two spins in the panel are up and two are down, none of the neighboring spins are convinced by the panel. 

The above rule is a modification of rule Ia of a two-dimensional Sznajd model introduced in \cite{sso00}. Stauffer et. al. assumed that a panel convinces its neighbors only if all four spins in the panel have the same direction. 

In all simulations we took initially $c_0$ of up-spins and $1-c_0$ of down-spins randomly distributed on the $L \times L$ lattice. This corresponds to a situation when the initial concentration of customers of product A is $c_0$. 
If the above rule was the only one imposed on the market then a phase transition would be observed: for $c_0<0.5$, after some relaxation time, the steady state with all down-spins (product B conquers the market) would be reached, whereas $c_0>0.5$ would lead to an all up-spin state (product A conquers the market). In case of $c_0=0.5$ each steady state is reached with probability $1/2$. This is in accordance with the results of Stauffer et. al. \cite{sso00}, which were obtained for a "weaker" rule.

Now we are in the position to introduce advertising into our model. Suppose that the concentration $c_0$ of buyers of product A ($\uparrow$) is less than 0.5. The question that immediately pops up is: "How strong should the advertising of product A be to conquer the market?" In the language of our model "conquering the market by product A" is equivalent to reaching a ferromagnetic steady state with all up-spins. 

To answer this question we introduce a kind of an external field $h \in [0,1]$, which may be treated as a measure of the level of advertising in the market (see also \cite{ps02,schulze03} where advertising was introduced in a similar way). We assume that if a customer is not convinced by its neighbors he/she will be more responsive to advertising and with probability $h$ will choose product A ($\uparrow$), see Fig. 2. Note that the introduction of the advertising rule does not change the steady states, i.e. the system always reaches a ferromagnetic steady state (all up-spins or all down-spins).

\begin{figure}[htbp]
\centerline{(a) \epsfxsize=3cm \epsfbox{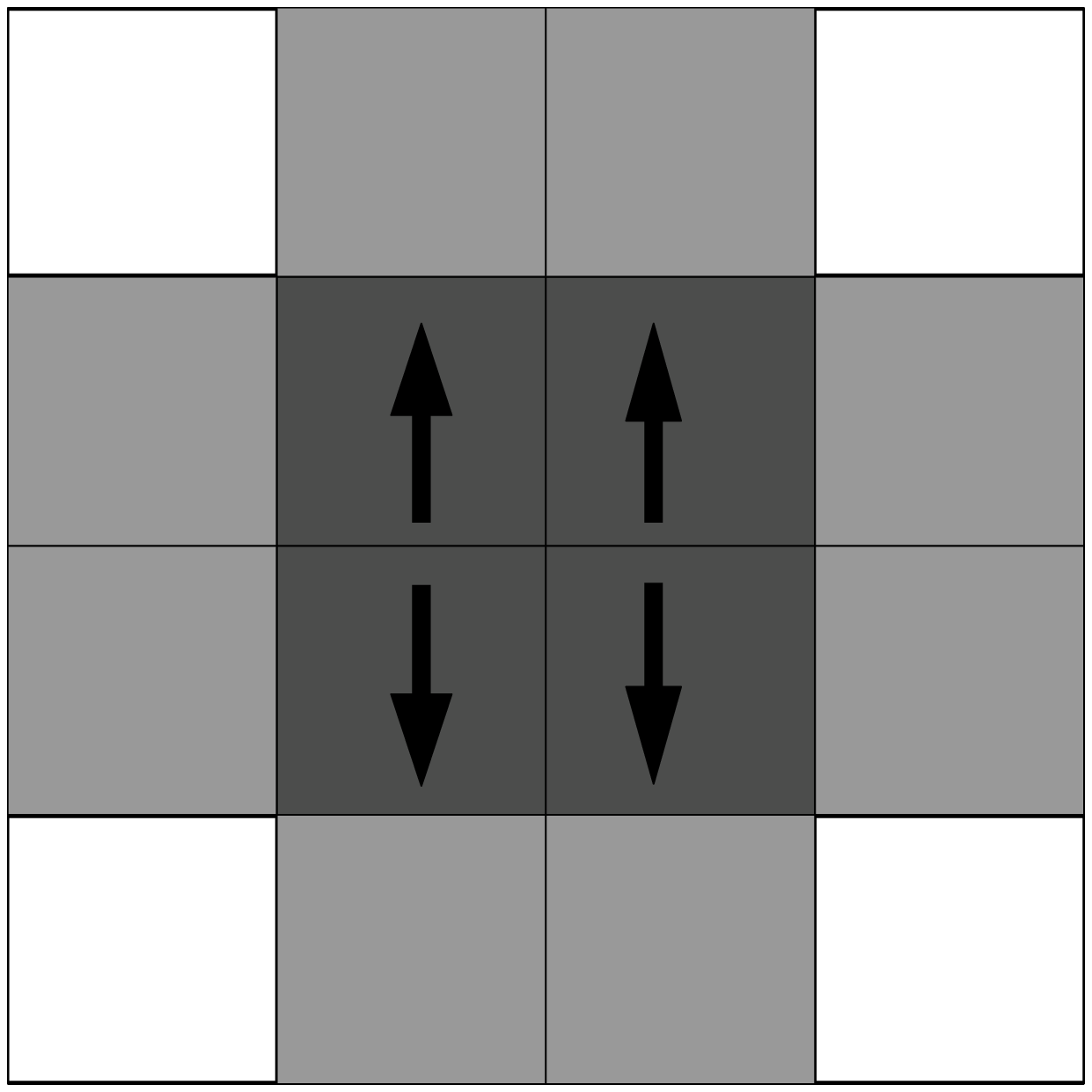}\qquad
            (b) \epsfxsize=3cm \epsfbox{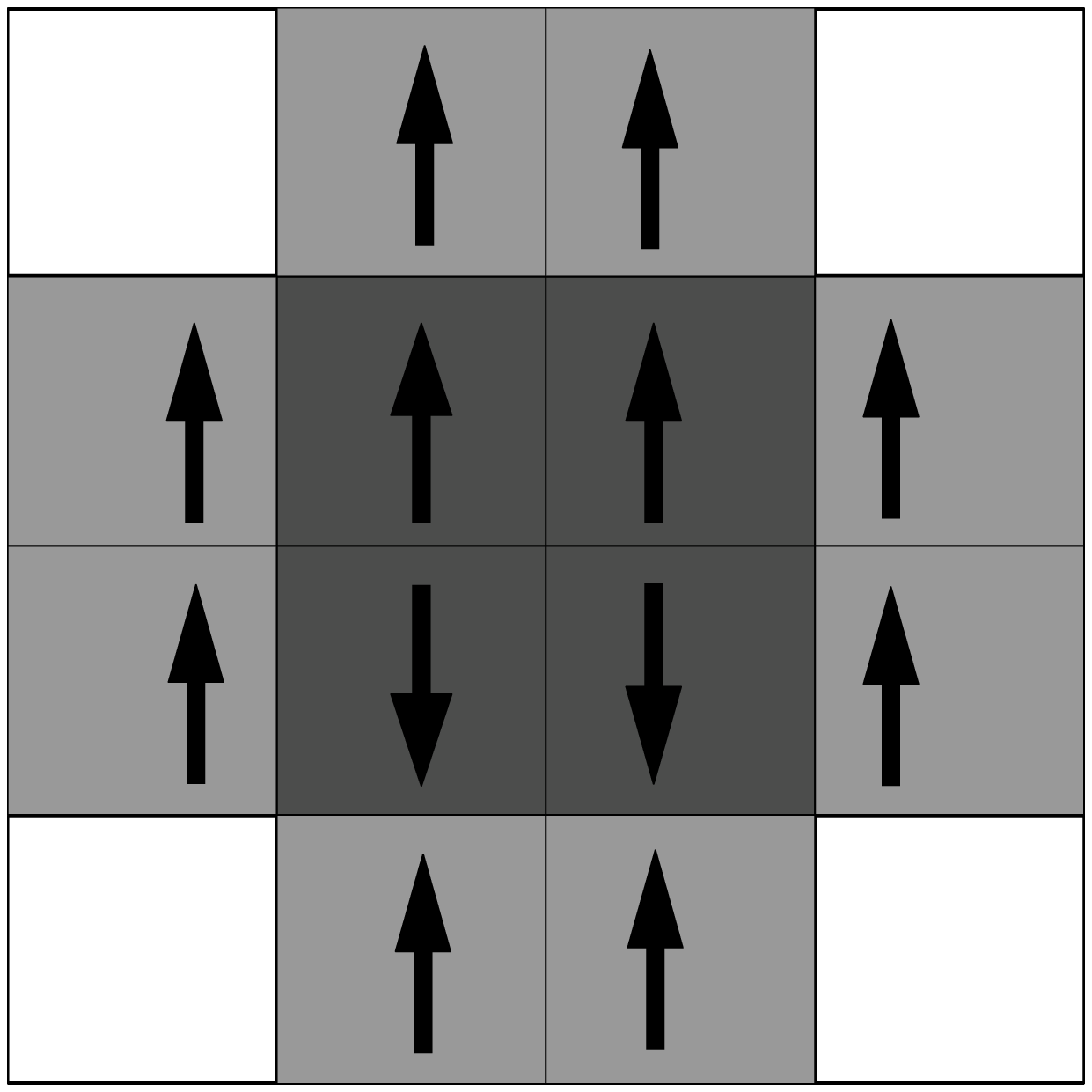}}             
\centerline{(c) \epsfxsize=3cm \epsfbox{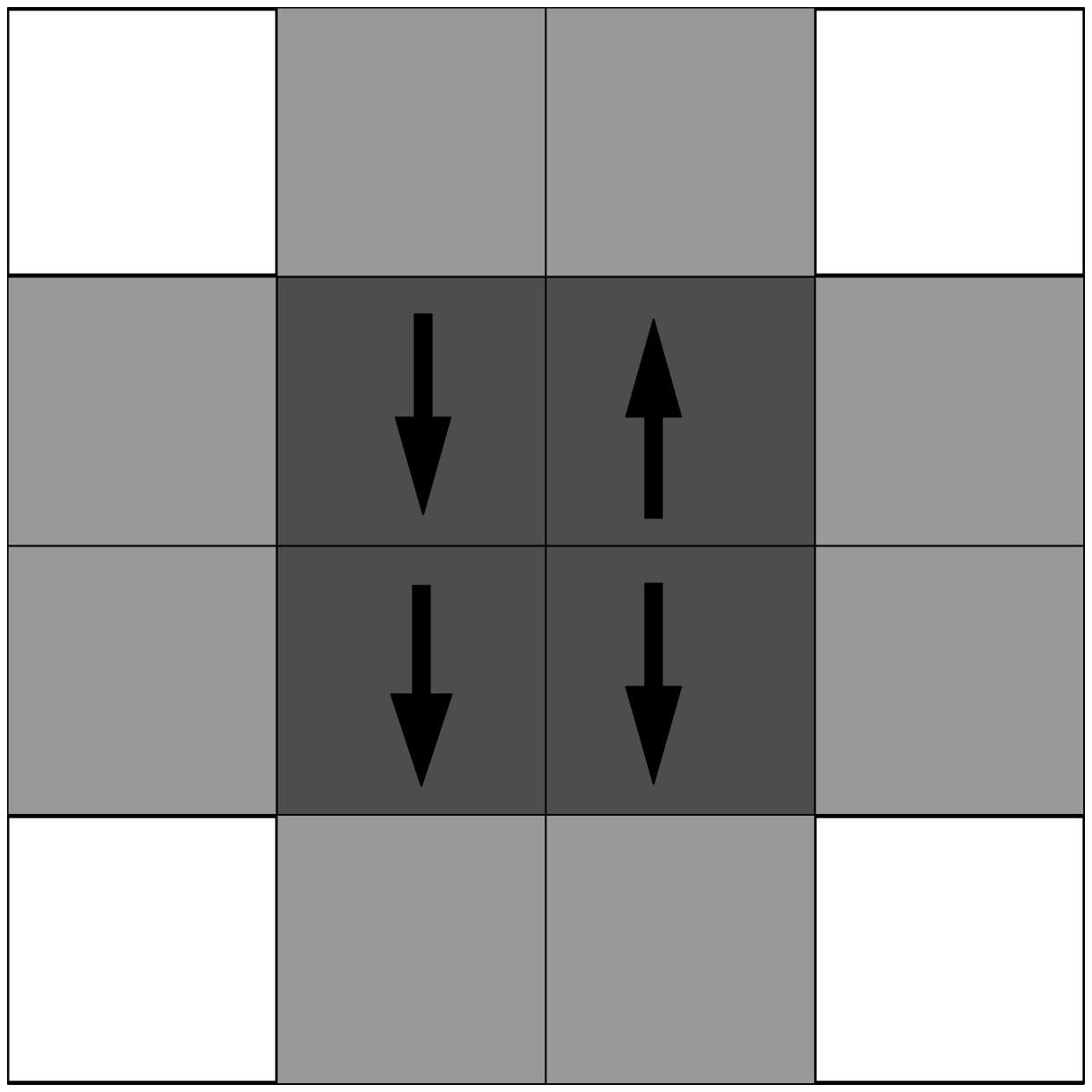}\qquad 
            (d) \epsfxsize=3cm \epsfbox{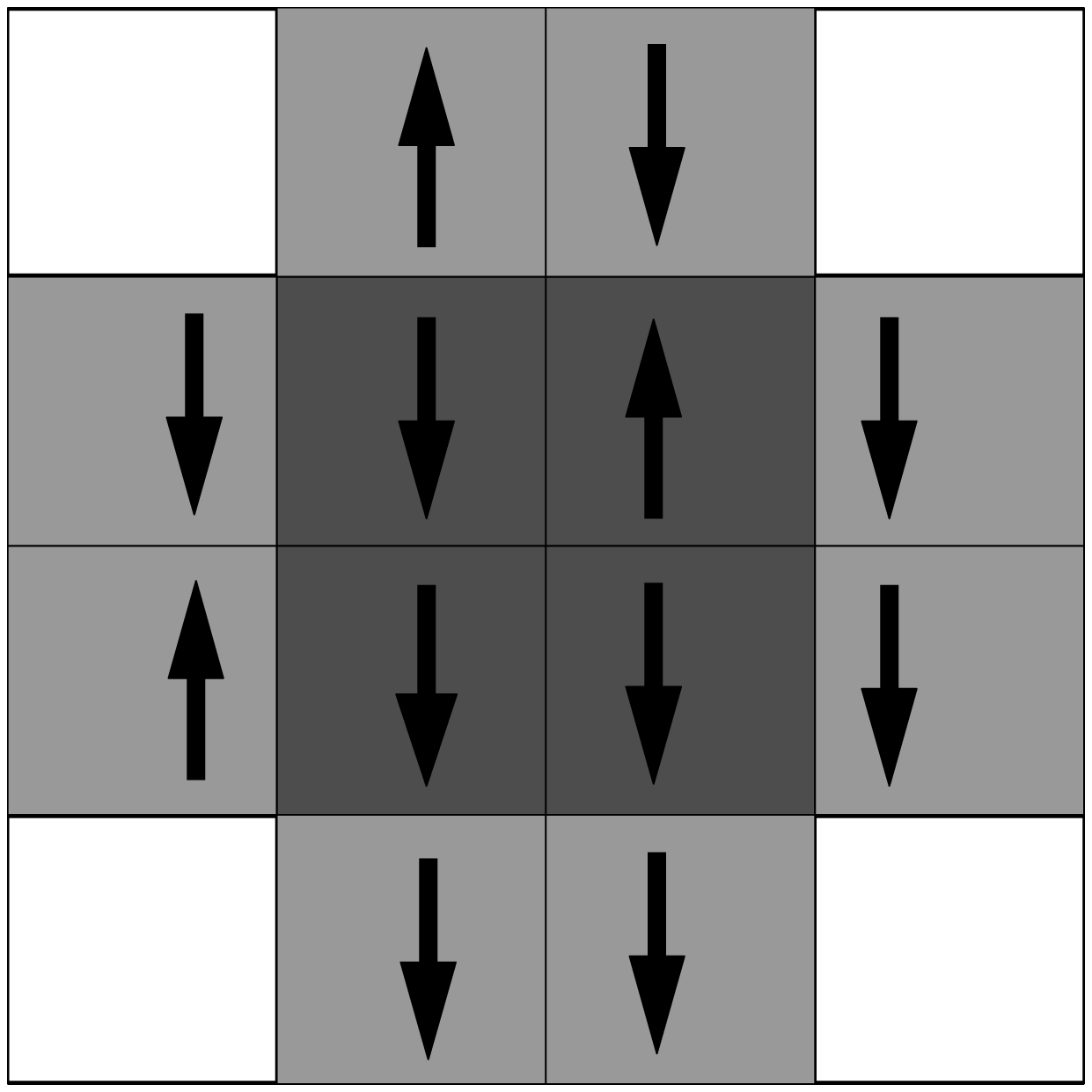}} 
\caption{
(a) If a panel consists of two customers of product A (up-spins) and two customers of product B (down-spins) then (b) all eight neighbors will be responsive to advertising. In the case of very strong advertising $h=1$ all of them will choose product A.
(c) If a panel consists of one customer of product A (up-spin) and three customers of product B (down-spins) then (d) with probability $3/4$ neighbors will choose product B (which is preferred by the majority) and with probability $1/4$ will be responsive to advertising. In the case of very strong advertising $h=1$ all responsive neighbors will choose the advertised product, i.e. product A.}
\end{figure}

\section{Results}

To investigate our model we performed standard Monte Carlo simulations with random sequential updating. We considered a square lattice of $L \times L$ Ising spins (i.e. up-arrows or down-arrows) with periodic boundary conditions. For most of the presented results $L=101$, however, we performed simulations for $L=31,53,71,307$ as well. We were usually making averaging over $10^3$ samples, with the exception of the smallest lattice where we averaged over $10^4$ samples. In all simulations we took initially $c_0$ of up-spins and $1-c_0$ of down-spins randomly distributed on the lattice. The results of the simulations are presented in Figs. 3-6. 

In Figure 3 we plotted the probability P($\uparrow$) of reaching the final steady state with all up-spins as a function of the initial concentration $c_0$ and the level of advertising $h$. Two kinds of regime changes (or phase transitions) can be observed. Firstly, for a given level of advertising $h$ there exists a "critical" value of the initial concentration $c_0^*$ of product A customers above which -- with probability one -- this product will conquer the market, i.e. the market will evolve to a steady state with all up-spins. Secondly, for an initial concentration $c_0$ of product A customers there exists a "critical" level of advertising $h^*$ of this product above which -- with probability one -- product A will conquer the market. Observe that there is an intermediate region in which the market evolves to the final up-spin state with probability $0<P(\uparrow)<1$, i.e. each of the two products has a possibility of conquering the market. For a very small level of advertising product A has no chance of winning and product B will always conquer the market, see Fig. 3.

\begin{figure}[htbp]
\centerline{\epsfxsize=8.5cm \epsfbox{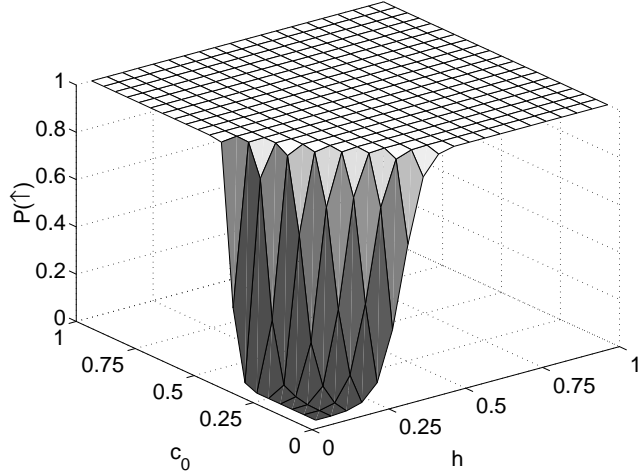}}
\caption{Probability P($\uparrow$) of reaching the final steady state with all up-spins as a function of the initial concentration $c_0$ and the level of advertising $h$.}
\end{figure}

In Figure 4 we presented the relaxation time $\tau$ needed to reach a steady state as a function of the level of advertising $h$ for several initial concentrations $c_0$ of product A customers. For the level of advertising slightly below the "critical" value $h^*$ the system relaxes very slowly, i.e. customers have a problem to decide which of the two products they should choose. Above the "critical" value $h^*$ the system reaches the all up-spin steady state much faster. This suggests that the level of advertising should be larger than $h^*$ in order to prevent the advertising counter-campaign. The relaxation time $\tau$ depends not only on the level of advertising $h$ but also on the initial concentration $c_0$ of product A customers. This is presented in Fig. 5. 

\begin{figure}[htbp]
\centerline{\epsfxsize=8.5cm \epsfbox{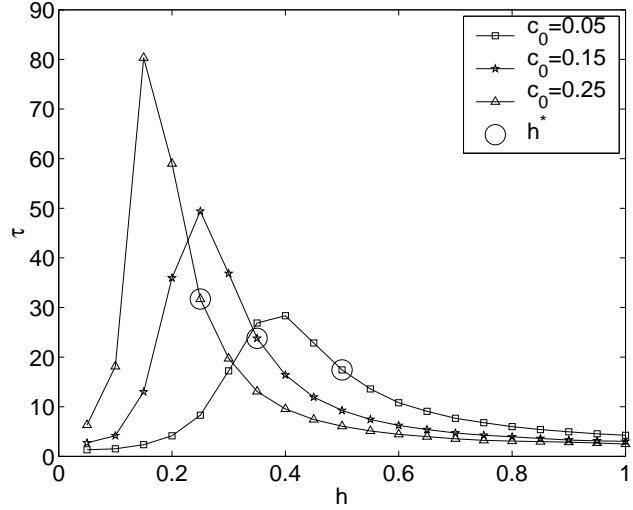}}
\caption{The relaxation time $\tau$ as a function of the level of advertising $h$ for several initial concentrations $c_0$ of product A customers. For the level of advertising slightly below the "critical" value $h^*$ the system relaxes very slowly.}
\end{figure}

\begin{figure}[htbp]
\centerline{\epsfxsize=8.5cm \epsfbox{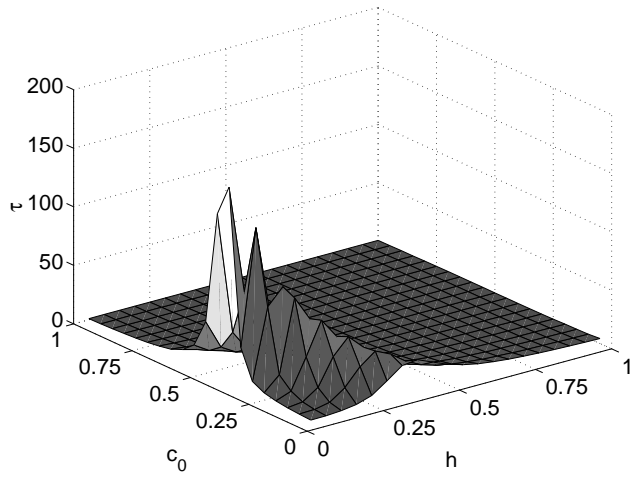}}
\caption{The relaxation time $\tau$ as a function of the initial concentration $c_0$ of product A customers and the level of advertising $h$.}
\end{figure}

The critical value of advertising $h^*$ as well as the size of the intermediate region depends clearly on lattice size. The larger the system the lower is the "critical" level $h^*$ above which product A conquers the whole market and the steeper is the regime change. For an infinite size lattice we may expect that there is no intermediate region -- below $h^*_\infty$ product B conquers the market with probability one and above $h^*_\infty$ product A conquers the market with probability one. Moreover, the "critical" level can be approximated. For example, $h^*_\infty \approx 0.28$ for the initial concentration of product A customers $c_0=0.05$, see Fig. 6. For other values of $c_0$ similar behavior is observed.

\begin{figure}[htbp]
\centerline{\epsfxsize=8.5cm \epsfbox{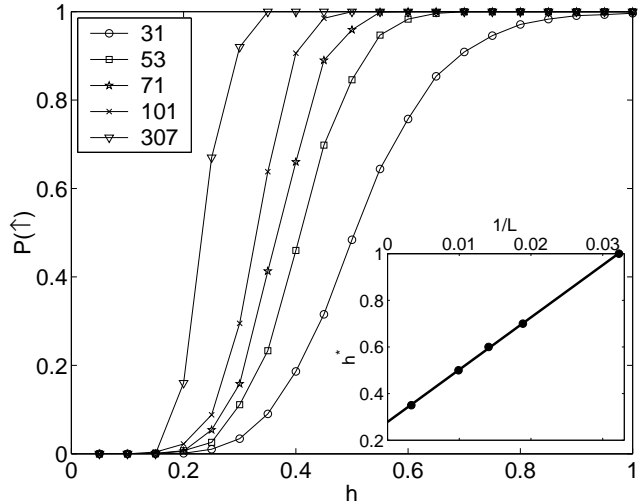}}
\caption{Dependence between the probability $P(\uparrow)$ that product A conquers the market and the level of advertising $h$ for five different lattice sizes $L=31,53,71,101,307$. The inset presents the dependence between lattice size $L$ and the "critical" level of advertising $h^*$. The initial concentration of product A customers is $c_0=0.05$.}
\end{figure}

\section{Conclusions}

In this paper we were studying the following situation: initially concentration of product A customers is $c_0$ and concentration of product B customers is $1-c_0$. The choice a customer makes (between two products sold in a duopoly market) is influenced by the opinion of his/her neighbors and advertising. If the neighboring panel consists of four individuals buying one product, the customer will buy the same one. If three individuals in the neighboring panel favor one product, the customer will buy this product with probability $3/4$. If the customer does not decide to follow the majority's opinion he/she will be responsive to advertising. The response of the customer to advertising (in this model only product A is being advertised) is measured by the parameter $h$. Clearly, the stronger the advertising, i.e. the higher the level $h$, the larger is the probability that product A will conquer the market. However, there exists a "critical" level of advertising $h^*$ above which product A will conquer the market with probability one. This suggests that in case when the manufacturer of product B does not react to advertising in a certain, limited amount of time it is not efficient for the seller of product A to make the advertising campaign more intense.

In real world oligopoly markets, innovation in terms of advertising and product introduction has been critical in keeping the brand image contemporary. However, the opportunity to create a sustainable competitive advantage through such actions is limited in many markets because of the ease with which competitors can replicate the strategy \cite{nn98}.
In our model we have found that above the "critical" value $h^*$ the relaxation of the system is rather fast preventing an advertising counter-campaign. On the contrary, if the level of advertising is close, but lower than $h^*$ the customers need much longer times to make a final choice between the two products, which makes the advertising campaign more costly and less efficient. We also have found that in larger systems the advertising level needed to conquer the market is smaller, which is in agreement with the results obtained by Schulze \cite{schulze03}. 

Naturally, calibrating the model to market data and finding the "critical" level of advertising is a story of its own. Our future research will be devoted to this subject.

\section{Acknowledgments}

The first author's research was partially supported by the Foundation for Polish Science (FNP) scholarship. The second author's research was partially supported by KBN Grant PBZ-KBN 016/P01/99.

\end{document}